\documentclass[conference]{IEEEtran}
\IEEEoverridecommandlockouts

\usepackage{cite}
\usepackage{amsmath,amssymb,amsfonts}
\usepackage{algorithmic}
\usepackage{graphicx}
\usepackage{textcomp}
\usepackage{xcolor}

\def\BibTeX{{\rm B\kern-.05em{\sc i\kern-.025em b}\kern-.08em
    T\kern-.1667em\lower.7ex\hbox{E}\kern-.125emX}}
    
\usepackage{booktabs}       
\usepackage{multirow}
\usepackage{threeparttable}
\usepackage{algorithmic}
\usepackage[ruled, vlined, linesnumbered]{algorithm2e}
\usepackage{subfigure}
\usepackage{array}
\usepackage{amsthm}
\usepackage{threeparttable}
\usepackage{url}
\usepackage{hyperref}
\usepackage[hyphenbreaks]{breakurl}

\begin{document}

\title{Predicting Protein-Ligand Binding Affinity via Joint Global-Local Interaction Modeling
}
\author{

    \IEEEauthorblockN{Yang Zhang$^{a,b*}$, Gengmo Zhou$^{a,b}$, Zhewei Wei$^{a}$, Hongteng Xu$^{a}$}
    \IEEEauthorblockA{$^a$ Renmin University of China}
    \IEEEauthorblockA{$^b$ DP Technology}
}

\maketitle

\begin{abstract}
The prediction of protein-ligand binding affinity is of great significance for discovering lead compounds in drug research. 
Facing this challenging task, most existing prediction methods rely on the topological and/or spatial structure of molecules and the local interactions while ignoring the multi-level inter-molecular interactions between proteins and ligands, which often lead to sub-optimal performance. 
To solve this issue, we propose a novel global-local interaction (GLI) framework to predict protein-ligand binding affinity. 
In particular, our GLI framework considers the inter-molecular interactions between proteins and ligands, which involve not only the high-energy short-range interactions between closed atoms but also the low-energy long-range interactions between non-bonded atoms. 
For each pair of protein and ligand, our GLI embeds the long-range interactions globally and aggregates local short-range interactions, respectively. 
Such a joint global-local interaction modeling strategy helps to improve prediction accuracy, and the whole framework is compatible with various neural network-based modules. 
Experiments demonstrate that our GLI framework outperforms state-of-the-art methods with simple neural network architectures and moderate computational costs. 
\end{abstract}

\begin{IEEEkeywords}
Protein-ligand binding affinity, graph neural networks, long-short interactions, drug discovery
\end{IEEEkeywords}

\section{Introduction}
In the research of drug discovery, protein plays a central role that affects specific functions in species and thus is treated as the target of drugs. 
Given an enormous amount of ligands, which refer to small molecular compounds having the potential to be drugs, we are interested in discovering the leading compounds whose bindings with target proteins affect the expressions of the proteins' functions and thus affect disease rehabilitation and our health. 
Therefore, the calculation of protein-ligand binding affinity is of great significance~\cite{2016Ligand,2004Docking, fu2021probabilistic} since the binding affinity reflects the binding strength and accordingly determines whether a ligand can be used as a candidate drug or not. 

In practice, the protein-ligand binding affinity can be measured in the laboratory, which, however, is very expensive and time-consuming~\cite{2004Docking,Trott2009AutoDock, tang2021artificial}. 
To reduce the cost and improve the efficiency of drug discovery, many methods have been proposed to predict the binding affinity with the help of computer science~\cite{Ballester2010A, 2015Improving}. 
Specifically, most existing methods can be coarsely categorized into two classes: simulation-based methods and machine learning-based methods. 
Simulation-based methods~\cite{2010Prediction, 2004Structural} apply expert knowledge and physical principles to simulate the binding process of proteins and ligands, whose computational efficiency and scalability are often questionable. 
The classic machine learning-based methods~\cite{Maciej0Development, 2015Improving, Guoping2012Performance} mainly depend on hand-crafted features like ECFP (extended connectivity fingerprint) of proteins and ligands, and leverage off-the-shelf techniques like random forest~\cite{2015Improving} to predict the binding affinity. 
These early works are fast, but their prediction accuracy is limited because of the loss of information caused by manual feature engineering. 
Recently, many new machine learning-based methods have been developed with the help of deep learning techniques, in which the features are learned from the molecular data rather than designed manually. 
In particular, various deep learning models are proposed based on convolutional neural networks (CNNs)~\cite{2015AtomNet, 2019DeepConv, 2019OnionNet, 2021OnionNet} and graph neural networks (GNNs)~\cite{2017graph2vec, 2017TopologyNet, 2018How, 0Development}, which outperform classic learning-based methods consistently. 

\begin{figure}[t]
	\centering
	\includegraphics[width=\linewidth]{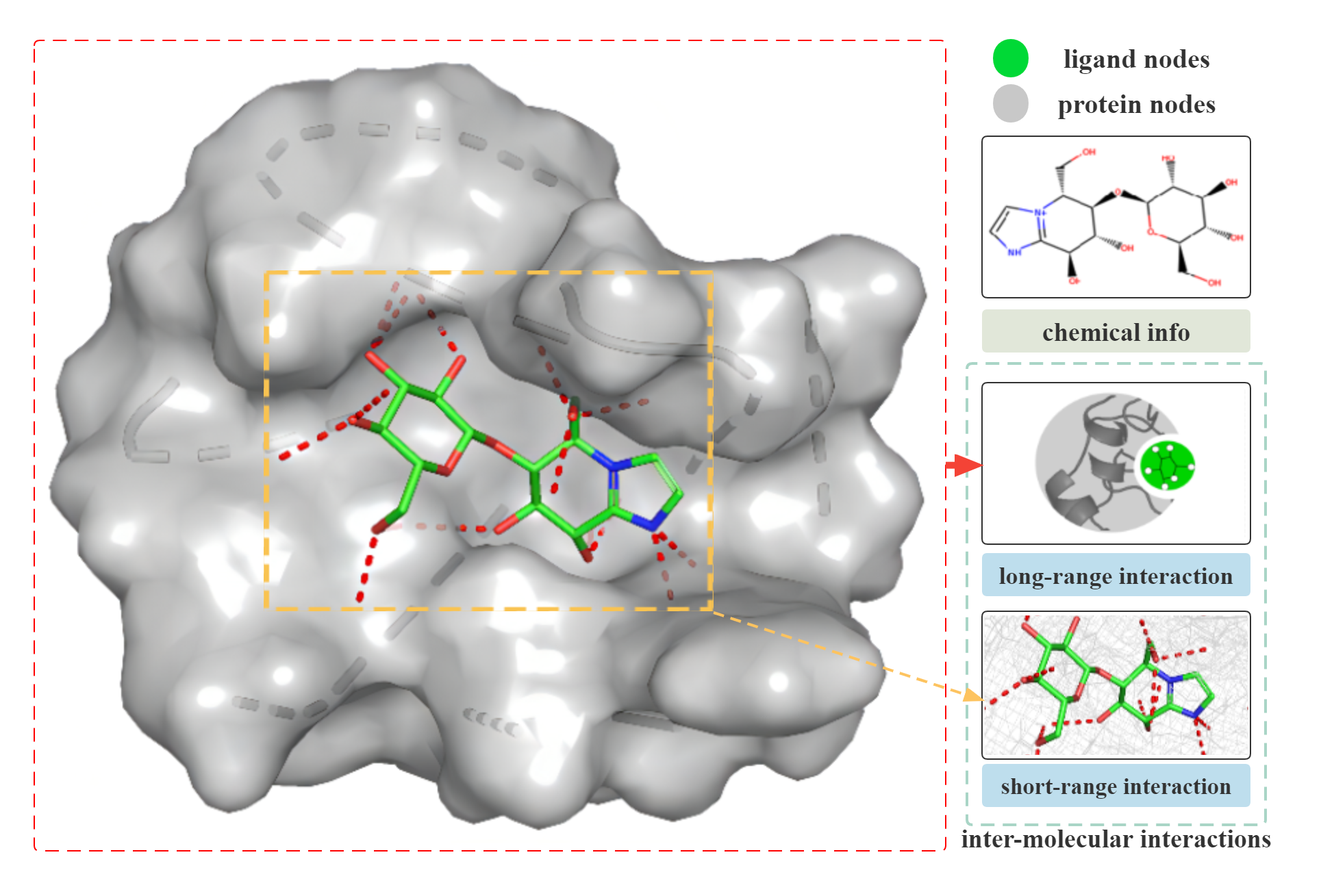}
	\caption{An illustration of the multi-level interactions between a protein and a ligand.}
	\label{fig:interaction}
\end{figure}

Although the above methods have achieved encouraging progress in predicting protein-ligand binding affinity, they still suffer from the following drawbacks in their modeling principles. 
In particular, as shown in Figure~\ref{fig:interaction}, the binding affinity of a protein and a ligand is mainly determined by the inter-molecular interactions between them, which consist of \textit{long-range} and \textit{short-range} interactions~\cite{1980Long, 1992Long, 2009Theory, S2013Effective, 2021MoCL, Xuedong1997Calculation}, respectively. 
The long-range interactions come from non-bonded atoms that might be weak but have a large amount. 
In contrast, the short-range interactions correspond to the non-covalent interactions between the atoms close to each other, which have few amounts but own high energy in general. 
Such interactions are based on various factors, including the biochemical information of the protein and the ligand, their topological (2D) structure, their spatial (3D) structure, and so on. 
Facing such multi-level interactions and complicated structural information, existing methods, however, merely consider the short-range interactions and local non-spatial structural information when predicting the binding affinity, which inevitably leads to sub-optimal performance. 
Although some attempts have been made recently to leverage spatial~\cite{Sch2017SchNet,2020Directional,2019Predicting,liu2021spherical} and pairing information~\cite{li2021structureaware,2020InteractionNet} of proteins and ligands, they do not provide systematic solutions to model both long-range and short-range interactions jointly when predicting the binding affinity. 

To overcome the drawbacks above, we propose a novel global-local interaction (GLI) framework to predict protein-ligand binding affinity. 
In particular, the proposed framework is comprised of three modules:
\romannumeral1) A chemical embedding module is applied to model the intra-molecular interactions happening within proteins and ligands and embed the molecules accordingly, which is implemented based on graph neural networks. 
\romannumeral2) A global interaction module is used to learn the long-range interactions between proteins and ligands, which embeds the chemical and spatial information of proteins and that of ligands, respectively, and then models global interaction effects based on the embeddings. 
\romannumeral3) A local interaction module is used to learn the short-range interactions between proteins and ligands, which models the local interaction effects based on some paired atoms that own non-covalent bonds and have short distances. 
Finally, given a protein and a ligand, their binding affinity is predicted based on both global and local interaction effects derived from the above modules. 

The main contributions of this paper are as follows: 
\begin{itemize}
 \item To the best of our knowledge, we are among the first to develop a general algorithmic framework for protein-ligand binding affinity prediction, which captures the multi-level inter-interactions between proteins and ligands. 
 \item In particular, the proposed GLI framework models the long-range and short-range interactions in the binding process of proteins and ligands, which has strong explanatory ability, generalization ability, and great potential for expansion. 
 \item Our framework has high flexibility — the three modules can be designed separately and learned jointly, each of which is highly compatible with various neural network architectures.
 \item We implement our GLI framework based on some simple neural network architectures and evaluate it on the well-known PDBbind dataset~\cite{2004The}. 
Experimental results demonstrate the superiority of our GLI to the state-of-the-art methods in both prediction accuracy and efficiency. 
\end{itemize}

\section{Related Work}

\subsection{Conventional Methods}
Accurate prediction of protein-ligand binding affinity is crucial for drug screening and discovery. Researchers have tried a variety of approaches to overcome this problem. 
On the one hand, scientists have proposed simulation-based methods~\cite{2010Prediction, 2004Structural} (like Autodock vina~\cite{Trott2009AutoDock}). 
With the spatial structure of protein and ligand, these methods simulate the mutual movement of protein and ligand. They are based on physical and biological knowledge to predict the binding affinity.
However, as aforementioned, these methods are time-consuming and have limited accuracy. 
Furthermore, they are difficult to be improved. 
On the other hand, based on expert experience, scientists extract some hand-crafted features from proteins and ligands, like fingerprints~\cite{2014Does, Guoping2012Performance, Maciej0Development} (ECFP, etc.), which represent local structures of molecules as features and then train machine learning models~\cite{2017Molecular, Ballester2010A, 2015Improving} (like random forest, SVM, and regression models) to make predictions. 
However, these methods are limited by manual feature engineering and often have unsatisfactory accuracy and weak generalization ability.

\subsection{Recent Deep Learning-based Methods}
With the development of deep learning techniques, automatic feature engineering and high-quality models become accessible.
Many models based on deep neural networks are proposed for the field of drug discovery. 
With the idea of word2vec~\cite{2013Efficient} in NLP, researchers proposed mol2vec~\cite{Sabrina2018Mol2vec} and other models to extract information from the SMILES (Simplified molecular input line entry system)~\cite{Weininger1988SMILES} (1D representation of molecule structure). 
However, there is limited chemical information in SMILES. 
Besides leveraging SMILES, CNN models have achieved great success in the field of computer vision. 
Therefore, by treating the molecular as an image, some studies use CNN models~\cite{2015AtomNet, 2019OnionNet, 2019Graph, 2019DeepConv, 2020DeepGS, 2020Improved, 2021OnionNet} to learn the local spatial structure and short-range interaction in the molecule. However, these methods ignore the long-range interaction information between atoms, and important chemical information like atom types, chemical bonds contained in a molecule, which will inevitably lead to loss. 
In order to maximize the extraction of chemical information, many researchers regard the atoms and bonds of molecules as topological structures and apply GNN models (like graph2vec~\cite{2016Learning}, GAT~\cite{2017Graph}, GCN~\cite{2016Semi}, GIN~\cite{2018How} and so on~\cite{ 2017TopologyNet, 2019Graph}) to extract the chemical information, which achieves good results. 
However, proteins and ligands have their own unique 3D spatial structure, such as angle and distance, which contains comprehensive and rich information for biological characterization. 
Moreover, with the improvement of biological structure algorithm~\cite{0}, the acquisition of protein structure information has become simpler than before. 
Therefore, on the basis of graph models, 3D spatial graph models (DimeNet~\cite{2020Directional} , DimeNet++~\cite{2020Fast}, SphereNet~\cite{liu2021spherical}, SchNet~\cite{Sch2017SchNet} and etc~\cite{2019Predicting}) aggregate spatial and topological information, which is not suitable for macromolecules (which consist of $10^3-10^7$ atoms) due to the large amount of calculation and resource consumption. 
Besides, from the perspective of research objects, most studies focus on the change of atom embedding in binding~\cite{2015AtomNet, 0Development, 2019OnionNet, 2019DeepConv, 2020Improved, 2020DeepGS, 2021OnionNet}, and predict the binding affinity by summarizing the processed atom embedding. 

The above methods have some flaws. Because, in essence, the binding affinity is determined by the inter-molecular interactions between the atoms of ligands and proteins rather than directly determined by atomic information.
Recently, a few researchers have also paid attention to this problem~\cite{li2021structureaware, 2020InteractionNet}, but these studies do not take the multi-level interactions between proteins and ligands into account. 
To solve the above problem, we propose a GLI model framework, which leverages both long-range and short-range interactions between proteins and ligands to predict the protein-ligand binding affinity. 
\section{Preliminaries}

\textbf{Protein and Ligand Graphs.}
We define a protein and a ligand as two graphs, denoted as $\mathcal{G}^{P}=(V^{P}, E^P)$ and $\mathcal{G}^{L}=(V^{L}, E^L)$, respectively.
For each graph $G$, we denote its nodes as a set $V=\{a_i\}$, where $a_i$ is the i-th atom node, and its edges as a set $E=\{e_{ij}\}$, where $e_{ij}$ is the edge connecting $a_{i}$ to $a_{j}$.

The proteins and ligands are attributed, which contain significant chemical and spatial information. 
Accordingly, each $\mathcal{G}^{P}$ (or $\mathcal{G}^{L}$) can be instantiated in the following two views:

\paragraph{Chemical Info Graph $\mathcal{G}_{c}=(V_{c}, E_c)$}
$V_{c}$ contains the chemical information of nodes, including their atom types, formal charges, chiral tags, aromatic tags, ring types (indicating whether a node in a ring or not), and degrees. 
$E_c$ contains information of chemical bonds, including bond types and lengths.

\paragraph{3D Spatial Graph $\mathcal{G}_{s}=(V_{s}, E_s)$}
$V_{s}$ contains the 3D coordinates of nodes, and $E_s=\{d_{ij}\}$ contains the weights of edges, where $d_{ij}$ represents the distance between $a_{i}$ and $a_{j}$. 
The edges in $E_s$ correspond to the node pairs whose distances are smaller than the cutoff distance $\mu$.



In this study, we use RDKIT~\cite{2014Bringing} to extract the biochemical and spatial information of proteins and ligands and construct $\mathcal{G}_c^P$, $\mathcal{G}_s^P$, $\mathcal{G}_c^L$, and $\mathcal{G}_s^L$ accordingly.
Furthermore, given a protein and a ligand, we consider the interactions between them and construct a local interaction graph as follows:

\paragraph{Local Interaction Graph $\mathcal{G}^{local}(V^{local},E^{local})$}
For each edge in $\mathcal{G}^{local}$, the corresponding atoms are from a protein and a ligand, respectively. 
The distance between the atoms is smaller than the cutoff distance $\mu$, and thus, there is a non-covalent interaction occurring between the atoms. 
Algorithm~\ref{alg:3} describes in detail the construction of the local interaction graph.

\begin{algorithm}[t]
	\renewcommand{\algorithmicrequire}{\textbf{Input:}}
	\renewcommand{\algorithmicensure}{\textbf{Output:}}
	\caption{Local Interaction Graph Construction}
	\label{alg:3}
	\begin{algorithmic}[1]
		\REQUIRE 
		The nodes sets $V^{P}, V^{L}$ and the spatial info sets $V^{P}_s, V^{L}_s$, the cutoff distance $\mu$
		\ENSURE The local interaction graph $\mathcal{G}^{local}(V^{local}, E^{local})$
		\STATE Initialize $V^{local} \gets \{\}, E^{local} \gets \{\}, $;
		\FORALL{atom nodes in $a_i \in V^L$ }  
		    \FORALL{atom nodes in $a_j \in V^P$ }  
		        \STATE Extract euclidean distance $d_{ij}$ between $a_i$ and $a_j$
		        \IF{$d_{ij} < \mu$}
		            \STATE Update $V^{local} \gets V^{local} \cup \{a_i,a_j\}$
		            \STATE Update $E^{local} \gets E^{local} \cup \{e_{ij}\} $
		      \ENDIF
		  \ENDFOR
		\ENDFOR
		\STATE \textbf{return} $G^{local}(V^{local}, E^{local})$
	\end{algorithmic}  
\end{algorithm}

\textbf{Problem Statement.}
Given a molecule like protein or ligand, we can construct protein graph $\mathcal{G}^{P}(V^{P}, E^{P})$, ligand graph $\mathcal{G}^{L}(V^{L}, E^{L})$ and $\mathcal{G}^{local}(V^{local}, E^{local})$ with cutoff distance $\mu$. Our goal is to construct a model $f(\mathcal{G}^{P},  \mathcal{G}^{L}, \mathcal{G}^{local}, \mu)$ to capture the chemical info, long-range interactions and short-range interactions in the binding process of protein and ligand, so as to predict the binding affinity accurately.


\section{Model Framework}

\begin{figure*}[t]
	\centering
	\includegraphics[width=1\linewidth]{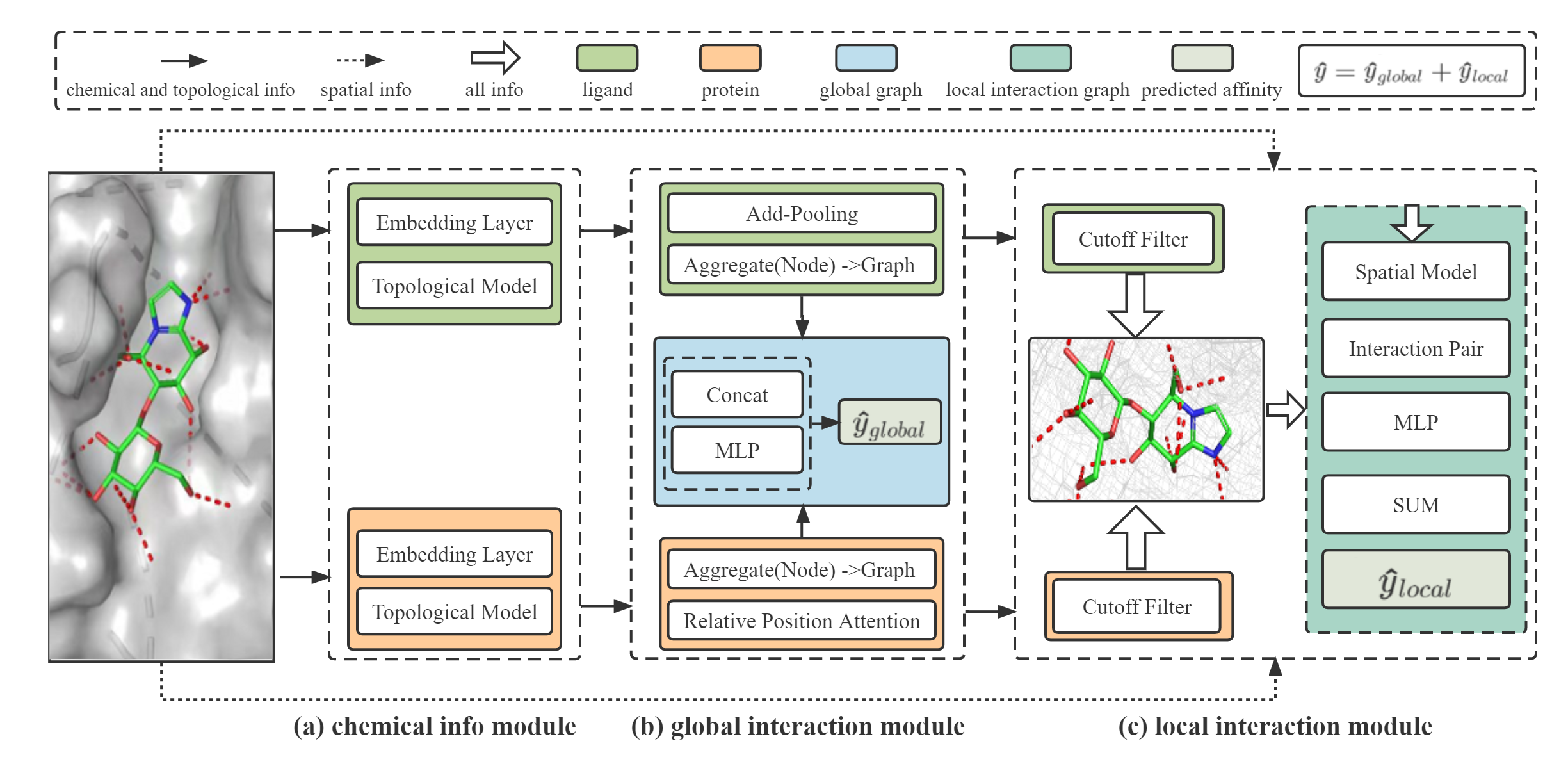}
	\caption{An illustration of our global-local interaction framework.}
	\label{fig:model}
\end{figure*}

Figure~\ref{fig:model} depicts the overall framework, which is composed of three modules: the chemical information module to gather the chemical information of atoms and chemical bonds; the global interaction module to represent long-range interactions; and the local interaction module to represent short-range interactions. 
The predicted binding affinity comes from the long- and short-range interactions. 

\subsection{Chemical Info Module}
Given a protein and a ligand, i.e., $\{\mathcal{G}_c^P,\mathcal{G}_s^P\}$ and $\{\mathcal{G}_c^L,\mathcal{G}_s^L\}$, we first apply a graph neural network to extract each atom's chemical embedding:
\begin{equation}\label{eq:1}
\mathcal{C}^P = f_{chem}(V_c^P\|V_s^P, E_c^{P}),~
\mathcal{C}^L = f_{chem}(V_c^L\|V_s^L, E_c^{L}),
\end{equation}
where $f_{chem}(\cdot)$ is the proposed chemical embedding model and ``$\|$'' represents the concatenation operation. 
It consists of two sub-modules. 
The first sub-module is an embedding layer, transforming chemical information to initial node embeddings and concatenating it to the corresponding spatial attribute;
Taking the concatenated node embddings and the topological information (i.e., $E_{c}^P$ and $E_c^L$) as input, the second sub-module (implemented via GAT, GCN, or GIN) extracts the chemical information embedding set of the protein and that of the ligand, denoted as $\mathcal{C}^P$ and $\mathcal{C}^L$, respectively.
Here, each $\mathcal{C}$ means a set of $c_i$'s, and $c_i$ is the chemical information embedding of the atom $a_i$. 

\subsection{Global Interaction Module}
Long-range interactions between non-bonded atoms are important components of inter-molecular interactions. 
Therefore, we presented a global interaction module for learning and representing long-range interactions between protein atoms and ligand atoms. 
Additionally, since it is extremely time-consuming and tedious to compute one atom by one atom, the module gathers the atomic level chemical embedding of ligand and protein as one node, respectively. 
Then the module estimates the interaction effect between overall protein $g^P$ and overall ligand $g^L$ as a long-range interaction $y_{global}$. 

\begin{figure}[t]
	\centering
	\includegraphics[width=1\linewidth]{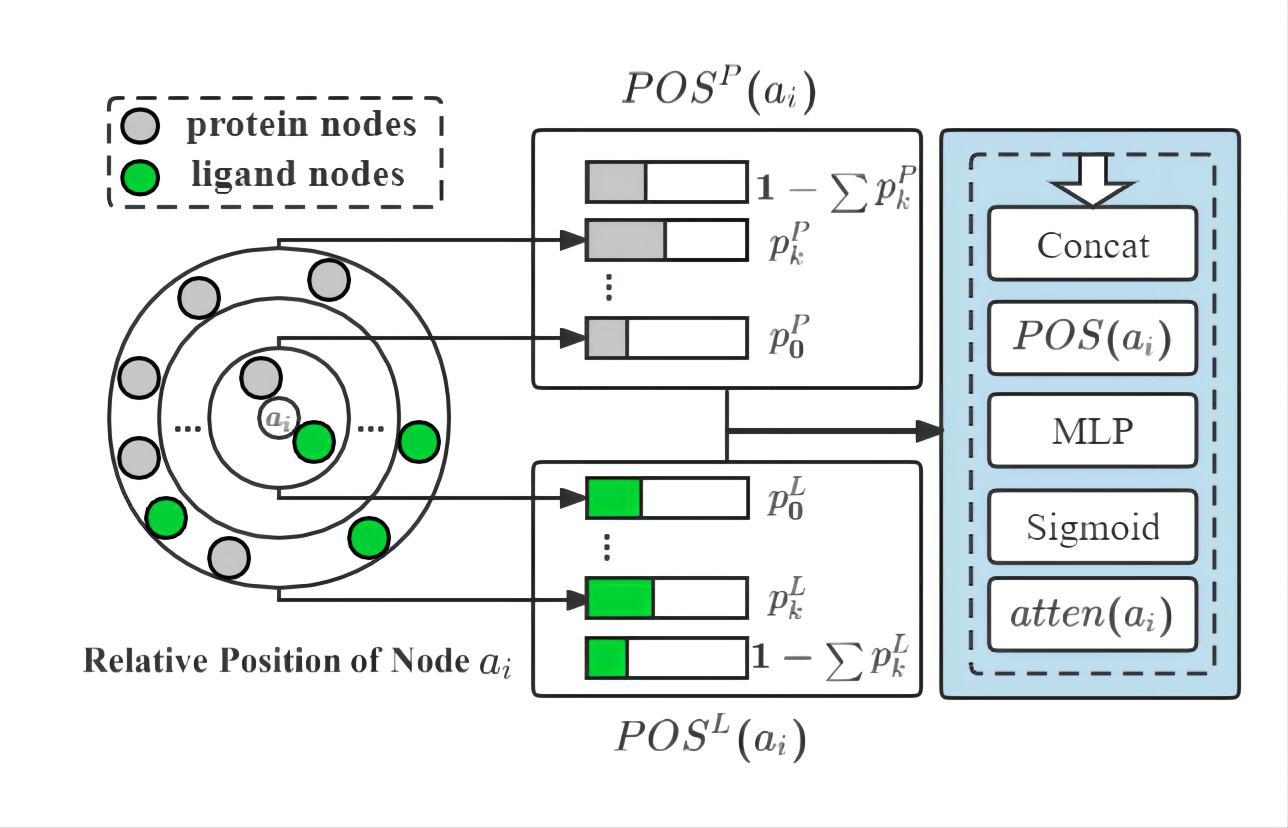}
	\caption{An illustration of Relative Position Attention.}
	\label{fig:position}
\end{figure}

\textbf{Relative Position Attention.}
In the calculation of long-range interactions, the position of the atom has some impact on the contribution of the interaction between atoms. 
For example, atoms' amounts and spatial distribution in proteins are far greater than those of ligands. 
In a protein, atoms close to the ligand should contribute more to inter-interactions than atoms far from the ligand. 
By referring to attention-pooling work~\cite{2015Gated, 2020Pushing}, we develop a new method named "Relative Position Attention" to express the importance of atoms in long-range interaction effects. 
As show in Figure~\ref{fig:position}, in particular, given an atom $a_i$, we derive two $K$-bin histograms for the remaining atoms in $\mathcal{G}^P$ and those in $\mathcal{G}^L$, respectively, according to their distance to $a_i$, i.e., 
\begin{equation}
\begin{aligned}
&POS^{P}(a_i)=[ p_{0}^{P}(a_i), ..., p_{K}^{P}(a_i), 1-\sideset{}{_{k=0}^{K}}\sum p_{k}^{P}(a_i)], \\
&POS^{L}(a_i)=[ p_{0}^{L}(a_i), ..., p_{K}^{L}(a_i), 1-\sideset{}{_{k=0}^{K}}\sum p_{k}^{L}(a_i)],
\end{aligned}
\end{equation}
where for $k=0,...,K$, 
\begin{equation}
\begin{aligned}
&p_{k}^{P}(a_i) = \frac{1(\{a_j|a_j \in \mathcal{G}^{P}\,\&\, k \le d_{ij} \leq k+1\})}{1(\{a_j|a_j \in \mathcal{G}^{P}\})}, \\
&p_{k}^{L}(a_i) = \frac{1(\{a_j|a_j \in \mathcal{G}^{L}\,\&\, k \le d_{ij} \leq k+1\})}{1(\{a_j|a_j \in \mathcal{G}^{L}\})},
\end{aligned}
\end{equation}
where $1(\mathcal{S})$ represents the cardinality of a set $\mathcal{S}$. 
These two histograms contain the relative position information of $a_i$, from the protein and the ligand, respectively. 
Accordingly, we leverage the following relative position attention mechanism to aggregate all such relative position information, i.e.,
\begin{equation}
POS(a_i)=[POS^{P}(a_i)\| POS^{L}(a_i)],
\end{equation}
\begin{equation}
atten(a_i)=\frac{\exp([c_i\| POS(a_i)])}{1 + \exp([c_i\| POS(a_i)])},
\end{equation}
\begin{equation}
g^P=\sideset{}{_{a_i\in \mathcal{G}^{P} }}\sum c_i \cdot atten(a_i),
\end{equation}
where $atten(a_i)$ represents the position attention, $c_i$ represents the chemical embedding of atom $a_i$ calculated in chem info module, and $K$ is a hyper-parameter that is set as 10 in this experiment.
$g^P$ is the comprehensive chemical information of the protein considering the position effect. 

In contrast, the ligand is much smaller and more concentrated in space. 
In order to reduce the computational cost, we use the add pooling method to aggregate the chemical information of the ligand $g^L$. 
\begin{equation}\label{eq:2}
g^L = \text{add-pooling}(\mathcal{C}^L), 
\end{equation}
At last, we concatenate the global embeddings of protein and ligand as the global interaction vector and apply a MLP (Multi-layer Perceptron) module to calculate the long-range interaction  $\hat{y}_{global}$. 
\begin{equation}\label{eq:3}
\hat{y}_{global} = \text{MLP}(g^P\|g^L),
\end{equation}

\subsection{Local Interaction Module}
The atoms within a range defined by the cutoff distance $\mu$ will have strong short-range interactions during the binding process. 
Our model contains a local interaction module to capture such information. 
As aforementioned, we first apply Algorithm~\ref{alg:3} to construct the local interaction graph from the protein and ligand according to the cutoff distance $\mu$. 
Then, to incorporate spatial information, we update the node embedding using the local spatial interaction model $f_{spatial}$. 
It can be set by returning the chemical embedding directly. 
It is also feasible to employ spatial graph-based models (like SchNet~\cite{Sch2017SchNet}, SphereNet~\cite{liu2021spherical}, DimeNet~\cite{2020Directional} and etc.) to update chemical and spatial embedding of all nodes $\mathcal{S}$ by incorporating the spatial information $V_s$. 
\begin{equation}\label{eq:6}
\mathcal{S}^{local} = f_{spatial}(\mathcal{C}^{local}, V^{local}_{s}),
\end{equation}
where $\mathcal{C}^{local}$ means the chemical info embedding set of the nodes in the local interaction graph $\mathcal{G}^{local}$, $V^{local}_{s}$ means the spatial info set of the nodes in local interaction graph $\mathcal{G}^{local}$.  
At last, we concatenate chemical and spatial node embedding $s_i, s_j$ of each non-covalent bond $e_{ij}$  between ligand nodes and protein nodes as the short-range interaction vector. 
Then, we feed all the vectors into an MLP and sum the results up as the prediction of the overall short-range interactions between protein and ligand. 
\begin{equation}\label{eq:7}
\begin{aligned}
\hat{y}_{local} = \sideset{}{_{(i,j)\in\{(i,j)|e_{ij}\in E^{local}\}}}\sum \text{MLP}(s_{i}||s_{j}).
\end{aligned}
\end{equation}

\subsection{Optimization Objective}
Since the binding affinity $y$ is determined by the long-range and short-range interactions between atoms of proteins and ligands, we estimate the binding affinity $\hat{y}$ by combining the results of the global interaction module $\hat{y}_{global}$ and the local interaction module $\hat{y}_{local}$. 
\begin{equation}\label{eq:8}
\hat{y} = \hat{y}_{global} + \hat{y}_{local}, 
\end{equation}
We use L1 loss function and the mean absolute error between the predicted binding affinity $\hat{y}$ and the measured ground truth $y$ to optimize the model. 
\begin{equation}\label{eq:9}
\mathcal{L} = \sideset{}{_{i=1}^{M}}\sum |y_i-\hat{y_i}|,
\end{equation}
where $\mathcal{L}$ represents the loss function, which is implemented as the mean absolute error (MAE), and $M$ is the number of protein-ligand pairs in the dataset. 
\section{Experiments}
In this section, we will conduct experiments on commonly used datasets to test the effects of our framework. 

\begin{table*}[tb] 
\footnotesize 
\begin{threeparttable}
\caption{Experimental results of baseline models and our GLI framework.\tnote{a}}
\label{tab:experiment result0}
\centering
\begin{tabular}{l c c c c c c c c c c c}
\toprule
 \multirow{2}{*}{Methods}     & \multirow{2}{*}{Type}  & \multicolumn{2}{c}{PDBbind v2016} & \multicolumn{2}{c}{PDBbind v2020} & \multicolumn{2}{c}{CSAR-HiQ} & GPU$\downarrow$ &  Time$\downarrow$ \\    
 \cmidrule(r){3-4}  \cmidrule(r){5-6}  \cmidrule(r){7-8} 
                              &                        &   MAE$\downarrow$           &     RMSE$\downarrow$   &   MAE$\downarrow$      &   RMSE$\downarrow$ &  MAE$\downarrow$    &  RMSE$\downarrow$    & (MB)  & (H) \\     \midrule
RF-Score         &                         Machine Learning     &   1.111(0.008)  &  1.397(0.008)     &  1.113(0.008)  & 1.398(0.004)     &  1.471(0.013)  &  1.847(0.015) &      -             &    0.08         \\ \midrule
Pafnucy    & \multirow{2}{*}{CNN}                           &   1.299(0.017)   & 1.597(0.019)  &   1.319(0.038) &   1.649(0.047)  &     1.482(0.051) & 1.853(0.050)  &       16,255      &        18.7            \\ 
OnionNet   &                            &   1.374(0.080)  & 1.813(0.113)     & 1.396(0.184)  & 1.815(0.274) & 1.355(0.032)   & 1.751(0.032)     &     30,701     &      0.25      \\ \midrule
GAT        & \multirow{8}*{Topological Graph}            & 1.307(0.016)  &    1.629(0.018)   & 1.294(0.013) & 1.606(0.014)  &  1.688(0.219) & 2.112(0.238)    &     1,403     &    \textbf{0.16}       \\ 
GCN        &                                             &  1.290(0.017)  &  1.597(0.021)    & 1.289(0.017) & 1.592(0.018)  &  1.620(0.316) & 2.041(0.349)   &    \textbf{1,243}      &     0.19       \\
GIN        &                                             &  1.249(0.022)    &  1.554(0.028) &  1.260(0.007) & 1.568(0.007)  &  1.442(0.093) & 1.849(0.104)      &      1,557 & 0.24 \\
GCN2       &                                             & 1.292(0.028)   &  1.604(0.032)  &   1.275(0.013) & 1.578(0.015)  &  1.872(0.128) & 2.316(0.135)    &  1,509  &0.24   \\
GAT + GCN  &                                             &  1.292(0.019)   & 1.599(0.023) &    1.273(0.011) & 1.593(0.014)  &  1.527(0.195) & 1.936(0.214)       &  1,385   &0.25      \\ 
GAT + GIN  &                                             &  1.241(0.014)   & 1.553(0.019)  &   1.246(0.019) & 1.559(0.023)  &  1.437(0.147) & 1.838(0.152) &         1,797   & 0.30           \\ 
GAT + GCN2 &                                             & 1.296(0.015)    &  1.601(0.015)   & 1.282(0.019) & 1.596(0.023)  &  1.705(0.114) & 2.128(0.127) &         1,835    & 0.45          \\ \midrule
SchNet     & \multirow{3}*{Spatial Graph}                & 1.157(0.035) & 1.454(0.035)  & 1.154(0.024) & 1.444(0.028)  &  1.330(0.064) & 1.715(0.066)     &    12,547       &    0.80              \\ 
DimeNet    &                                              &   -  &     - & -  &  -  & -  &  -   &      OOM     &     -        \\ 
SphereNet  &                                              &   -  &   -   & -  & -  & -  &   -  &        OOM     &      -       \\ \midrule
SIGN\tnote{b} &    Structure-Aware                      &  1.027(0.025)  &  1.316(0.031)  &  - &  - &                       1.327(0.040) & 1.735(0.031)   & 20,091  &  2.73        \\ \midrule
GLI-0      & \multirow{3}*{Global-Local}                &  1.049(0.021)  & 1.321(0.026)  & 1.060(0.019) & 1.354(0.020) &   1.230(0.028) & 1.595(0.039)   &  1,576  &  0.47  \\  
GLI-1      &                                            &  \textbf{1.026(0.023)}  & \textbf{1.294(0.034)}  & \textbf{1.059(0.023)} & \textbf{1.353(0.031)} &    1.232(0.043) & 1.597(0.046)   &  1,707  &  0.58     \\
GLI-2      &                                            &  1.078(0.023) & 1.347(0.032)   & 1.112(0.017) & 1.405(0.018)  &  \textbf{1.199(0.027)} & \textbf{1.549(0.032)}   &  1,635  &  0.45 \\ \bottomrule

\end{tabular}
  \begin{tablenotes} 
    \item[a] The standard deviation of each index is indicated in brackets. 
	\item[b] The result of SIGN comes from~\cite{li2021structureaware}, however, we cannot get similar results when utilizing public code and parameters.
 \end{tablenotes} 
\end{threeparttable}
\end{table*}

\begin{itemize}
    \item \textbf{Task 1.} Compared to the existing state-of-the-art models, is it possible for our framework to produce improved or even superior prediction results?
    \item \textbf{Task 2.} For different cutoff distances $\mu$, what will happen to the size of the local graph? Is the model result sensitive to this parameter?
    \item \textbf{Task 3.} Can different modules of our framework contribute to a substantial improvement? What contributions do the different modules make to the prediction, and are the results statistically significant?
    \item \textbf{Task 4.} How about the efficiency and computational consumption of our framework?
\end{itemize}

\subsection{Experiment Settings} \label{experiment}
\paragraph{Dataset} \label{dataset}
We will train and test the baseline models and our framework on the following datasets:
\begin{itemize}
\item \textbf{PDBbind v2016 }: In the related research of binding affinity prediction, PDBbind\footnote{http://www.pdbbind.org.cn/}~\cite{2004The} is a well-known and commonly used data set, which contains the 3D structures of proteins and ligands, binding pocket information, binding sites, and accurate binding affinity results determined in the laboratory. 
The data set consists of three sub-datasets (general set, refined set, and core set). 
The general set contains all protein-ligand complexes. 
The refined set contains better quality compounds selected from the general set. 
The core set includes carefully selected samples of the highest quality. 
In this experiment, consistent with other studies~\cite{2019OnionNet, li2021structureaware}, we will use the refined subset (4,057 samples) and core subset (290 samples) of the PDBbind v2016 to train and test our framework.
\item \textbf{PDBbind v2020}: This dataset is the latest version of the PDBbind dataset. 
It consists of two parts: general set(14, 127 complexes) and refined set(5,316 complexes).  
In our experiment, we train and test with the refined subset of the PDBbind v2020 and the core subset of the PDBbind v2016. 
\item \textbf{CSAR-HiQ\footnote{http://csardock.org/}}: This is another well-known dataset for predicting binding affinity, which contains two data sets with a total of 343 complexes~\cite{dunbar2011csar}. 
In our experiment, consistent with the study~\cite{li2021structureaware}, we use the refined subset of the PDBbind v2016 for training and take this data set for testing to verify the generalizability of our framework. 
\end{itemize}

\paragraph{Introduction to the binding affinity}
In drug chemistry experiments, binding affinity, such as $K_d$ (dissociation constant), $K_i$ (inhibition constant), or $IC_{50}$ (semi-inhibitory concentration), can describe the strength of ligand-protein binding. 
In the datasets of PDBbind and CSAR-HiQ, our target binding affinity is the negative logarithm of the experimental results (such as $-log K_d$, $-log K_i$, or $-log IC_{50}$).

\paragraph{Data Preparation}
We perform the following four data processing steps:
\romannumeral1) Remove the relevant sample data of the core subset from the refined subset in the PDBbind dataset. 
\romannumeral2) Use RDKIT \cite{2014Bringing} to read the conformation of protein, ligand and eliminate the samples with errors. 
\romannumeral3) PDBbind dataset contains protein and pocket files, and the atoms in the pocket are almost within 10 $\AA$ from the ligand. 
In contrast, CSAR-HiQ only has complex data (composed of proteins and ligands), and reading the entire protein data is both time-consuming and unnecessary.  
Therefore, to ensure the consistency of our experiment, we filter the atoms and edges in the protein within 10 $\AA$ from the ligand as pocket data. 
\romannumeral4) Extract the node and chemical bond features of the molecule data. 

\paragraph{Evaluation Metrics}
In order to evaluate the performance of different models and the significance of different modules in our GLI framework, we use mean absolute error (MAE), root mean square error (RMSE) to measure the model performance. 
For efficiency, we collect the average GPU RAM usage (MB) and running time (training, validating, and testing) (hours) of each model in each dataset. 
\begin{equation}\label{eq:10}
R M S E=\sqrt{\frac{1}{n} \sideset{}{_{i=1}^{n}}\sum\left(\hat{y}_{i}-y_{i}\right)^{2}} 
\end{equation}
\begin{equation}\label{eq:11}
M A E=\frac{1}{n} \sideset{}{_{i=1}^{n}}\sum\left|\hat{y}_{i}-y_{i}\right| 
\end{equation}
$\hat{y}_{i}$ and $y_{i}$ represent the predicted and experimental labels for the i-th complex in the dataset, respectively, which is the binding affinity between protein and ligand. 
n is the number of the dataset. 

Besides, we also performed a two-sample T-test to analyze whether different modules in our framework can bring significant performance changes. 
A t-test is a type of inferential statistic used to determine if there is a significant difference between the means of two groups. 
Generally, for two random variables $X_{1}$, $X_{2}$, we consider the hypothesis $H_{0}: \mu_{1} = \mu_{2}$. The test statistic is constructed as follows: 
\begin{equation}\label{eq:13}
T=\frac{\hat{\mu}_{1}-\hat{\mu}_{2}}{\sqrt{\frac{\hat{\sigma}_{1}^{2}}{\left|S_{1}\right|}+\frac{\hat{\sigma}_{2}^{2}}{\left|S_{2}\right|}}} 
\end{equation}
where $\hat{\sigma}_{1}^{2}$ and $\hat{\sigma}_{2}^{2}$ are unbiased estimates of the variances $\sigma_{1}^{2}$ and $\sigma_{2}^{2}$ of $X_{1}$ and $X_{2}$, respectively. 
$\left|S_{1}\right|$ and $\left|S_{2}\right|$ denote the corresponding sample sizes. 
If the P-value of $T$ is smaller than the $\alpha/2$ or more than $1-\alpha/2$, hypothesis $H_{0}$ is rejected. 

\paragraph{Baseline Models and Parameters}
We compare the experiment results with the following models: 
\romannumeral1) \textbf{Traditional machine learning models}: RF-score (Random forest~\cite{2015Improving}).  
The number of decision trees in RF-score is set to 100, the max-depth of trees is set to 5, the maximum number of features is set to 3 and the minimum number of samples required to split is set to 10. 
\romannumeral2) \textbf{CNN models}: Pafnucy~\cite{0Development} and Onionnet~\cite{2019OnionNet}. 
For the CNN-based models, we set the channels of three-layer 3D convolutions for Pafnucy as 64, 128 and 256. 
For OnionNet, the number of input features is 3840 and there are 32, 64, and 128 filters in the three convolutional layers with the kernel size of 4. 
\romannumeral3) \textbf{Topological (2D) graph-based models}:  GAT~\cite{2017Graph}, GCN~\cite{2016Semi}, GIN~\cite{2018How} and GCN2~\cite{chen2020simple}. 
For GNN-based models, the number of layers is set to 3, except for GAT. For GAT, we set the number to 1. 
For GIN, we set the initial $\epsilon$ to 0 and make it trainable. 
For GCN2, we set the strength of the initial residual connection $\alpha$ to 0.5 and the strength of the identity mapping $\beta$ to 1. 
\romannumeral4) \textbf{Spatial (3D) graph-based models}: SchNet~\cite{Sch2017SchNet}, DimeNet~\cite{2020Directional}, SphereNet~\cite{liu2021spherical}. 
For SchNet, DimeNet, and SphereNet as baseline models, we set the cutoff distance $\mu$ to 5. 
For Dimenet as a local spatail model in our framework, we set the cutoff distance $\mu$ to 4. 
\romannumeral5) \textbf{Structure-aware interaction model}: SIGN~\cite{li2021structureaware}. 
For SIGN, we use the recommended parameters\footnote{https://github.com/PaddlePaddle/PaddleHelix}. 

\begin{figure*}[t]
\caption{The influence of cutoff $\mu(\AA)$ for the local interaction graph and experimental results}
\label{fig: cutoff}
\centering
\subfigure[The per of protein and ligand nodes contained in local interaction graph]{
\includegraphics[scale=0.25]{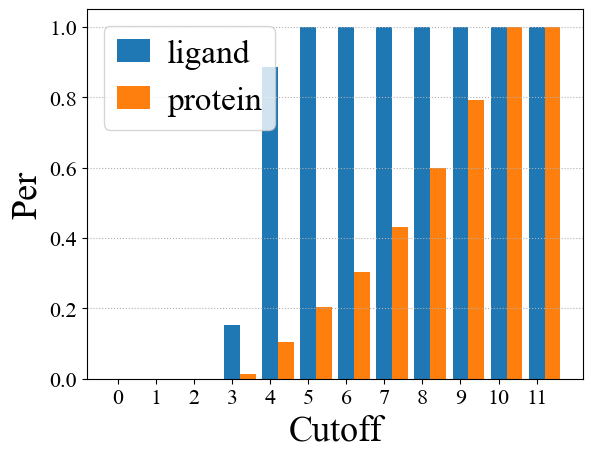} \label{1}
}
\subfigure[The number of protein and ligand nodes contained in local interaction graph]{
\includegraphics[scale=0.25]{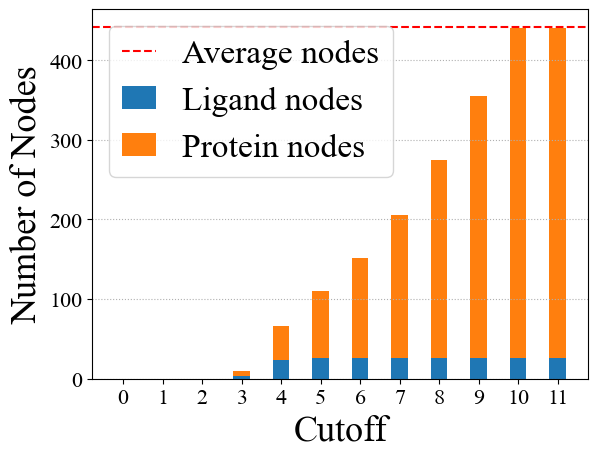} \label{2} 
}
\subfigure[The interaction edges between proteins and ligands in local interaction graph]{
\includegraphics[scale=0.25]{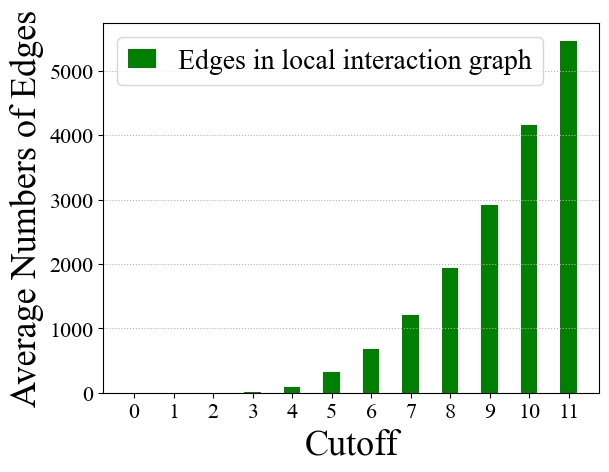}\label{3}
}
\subfigure[The test RMSE and MAE of GLI-1 with different cutoff in PDBbind v2016]{
\includegraphics[scale=0.25]{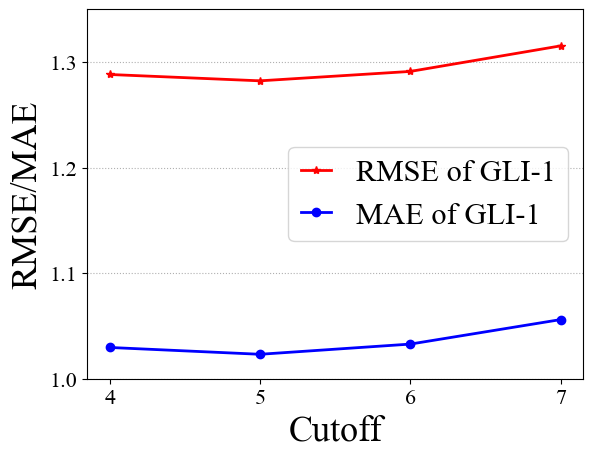}\label{4}
}
\end{figure*}

\paragraph{GLI Framework}
We implement our GLI framework based on different models in the following experiments. 
\begin{itemize}
\item  \textbf{GLI-0}: Taking (GAT + GCN) as the chemical info module. 
\item  \textbf{GLI-1}: Taking GIN as the chemical info module. 
\item  \textbf{GLI-2}: Taking GCN2 as the chemical info module. 
\end{itemize}

To indicate the GLI framework with different modules, we adopt the following symbols bellow:
\begin{itemize}
\item \textbf{GLI-*-c}: This means the GLI-* only contains the chemical info module. 
\item \textbf{GLI-*-cg}: Adding the global interaction module to the previous model. 
\item \textbf{GLI-*-cgl}: Adding the local interaction module to GLI-*-cg. 
\end{itemize}

\paragraph{Environment and Setings} \label{enviroment}
We implement our GLI framework in PyTorch and PyTorch Geometric, and all experiments are conducted on a machine with an NVIDIA V100 GPU (32GB RAM), Intel Xeon CPU (12 Cores, 2.5 GHz), and 92GB of RAM. 
We set the batch size as 32, dropout rate as 0.1 and use Adam optimizer for model training with a learning rate of 0.0001 and 200 epochs. 
We construct the local interaction graph with cutoff distance $\mu$ = 5 $\AA$ (the analysis about influences of $\mu$ for experiment shown in \textbf{Section~\ref{Cutoff analysis} }). 
The basic dimensions of node and edge embeddings are both set to 128. 
For baseline models and models in the chemical info module of GLI, we use suggested settings to get optimal performance. 
We use 10-fold cross-validation for training and validating. 
Related codes, processed data, and trained models for our experiments can be found at the anonymous link~\cite{codeurl}. 

\subsection{Result Analysis} 
\begin{table*}[tb]
\footnotesize
\begin{threeparttable}
\caption{Ablation results.\tnote{a}}
\setlength\tabcolsep{4pt} 
\label{tab:ablation result}
\centering
\begin{tabular}{l c c c c c c c c c c c}
\toprule
 Chem Info    & Global  & Local  & Local Spatial   & \multicolumn{2}{c}{PDBbind v2016} & \multicolumn{2}{c}{PDBbind v2020} & \multicolumn{2}{c}{CSAR-HiQ} & GPU$\downarrow$ &  Time$\downarrow$ \\    
 \cmidrule(r){5-6}  \cmidrule(r){7-8}   \cmidrule(r){9-10} 
     &     &  &      Model       &   MAE$\downarrow$        &     RMSE$\downarrow$   &   MAE$\downarrow$      &   RMSE$\downarrow$ &  MAE$\downarrow$    &  RMSE$\downarrow$    & (MB) & (H) \\     \midrule
GAT+GCN          &                                &        &     & 1.292(0.019)   & 1.599(0.023)          &    1.273(0.011) & 1.593(0.014)  &  1.527(0.195) & 1.936(0.214)       & \textbf{ 1,385}  &0.25      \\
GAT+GCN          &             \checkmark         &        &     &  1.111(0.028)    & 1.399(0.031)           & 1.146(0.016) & 1.457(0.022)  &  1.274(0.045) & 1.646(0.054)   &  1,479 & 0.31     \\ 
GAT+GCN          &             &        \checkmark         &     &  1.100(0.027)    &  1.409(0.032)         &  1.134(0.022) & 1.489(0.032)  &  1.429(0.053) & 1.811(0.064) &  1,559 & 0.43       \\ 
GAT+GCN          &               \checkmark       & \checkmark    &      &  1.049(0.021) & 1.321(0.026)   &  1.060(0.019) & 1.354(0.020)  &  1.230(0.028) & 1.595(0.039)   &  1,576      &  0.47   \\ 
GAT+GCN          &              \checkmark        &    \checkmark        &  SchNet    &  1.034(0.030)  & 1.314(0.039)   & 1.077(0.021) & 1.387(0.026)  &   1.203(0.035) & 1.555(0.042)   & 6,150  & 0.97       \\ 
GAT+GCN          &              \checkmark        &    \checkmark        &  DimeNet    &  1.123(0.038)    & 1.413(0.037)   &    1.152(0.021) & 1.472(0.031) & 1.964(0.477) & 2.419(0.534)      & 29,503  &2.55         \\ 
\midrule
GIN         &                    &                         &          &  1.249(0.022)   &  1.554(0.028)      &   1.260(0.007) & 1.568(0.007)   &   1.442(0.093) & 1.849(0.104)    &  1,557   & \textbf{0.24}        \\ 
GIN          &            \checkmark         &    &                     & 1.104(0.026)      & 1.392(0.026)   &   1.106(0.022) & 1.417(0.022)     & 1.260(0.034) & 1.638(0.038)  &     1,543    &       0.41      \\ 
GIN          &               &     \checkmark    &                     & 1.069(0.017)       & 1.375(0.021)   &   1.148(0.019) & 1.513(0.026)  &    1.422(0.045) & 1.825(0.052)   &      1,653    &       0.48     \\ 
GIN          &            \checkmark         &        \checkmark                &        &   \textbf{1.026(0.023)}   & \textbf{1.294(0.034)} & \textbf{1.059(0.023)} & \textbf{1.353(0.031)}    &  1.232(0.043) & 1.597(0.046)  &     1,707    &       0.58    \\ \midrule
GCN2          &                    &                         &          &1.292(0.028)    &  1.604(0.032)   &   1.275(0.013) & 1.578(0.015)  &  1.872(0.128) & 2.316(0.135)  &  1,509  & \textbf{0.24}          \\ 
GCN2          &            \checkmark         &               &        &  1.120(0.019)   &   1.407(0.016)  &   1.141(0.021) & 1.438(0.020)   & 1.224(0.033) & 1.587(0.034)    &   1,517      &         0.27   \\ 
GCN2          &               &   \checkmark          &           &  1.113(0.017)       &  1.415(0.024)  &     1.122(0.017) & 1.460(0.019) &   1.359(0.040) & 1.758(0.046) &  1,567      &          0.35   \\ 
GCN2          &            \checkmark         &        \checkmark                &          & 1.078(0.023) & 1.347(0.032)  & 1.112(0.017) & 1.405(0.018)   & \textbf{1.199(0.027)} & \textbf{1.549(0.032)}   &  1,635  &  0.45         \\ \bottomrule
\end{tabular}
  \begin{tablenotes} 
    \item[A] The standard deviation of each index is indicated in brackets. 
  \end{tablenotes}
 \end{threeparttable}
\end{table*}

\subsubsection{Model Comparison}
In Table~\ref{tab:experiment result0}, we compared the models based on our GLI framework with the five aforementioned types of baselines. 
The topological graph-based methods leverage the topological information of molecules, learning and aggregating the local structural features of proteins and ligands, e.g., atom types, chemical bonds and other biochemical information. 
The prediction performance of these models is inadequate, with the results of RMSE more than 1.5 in PDBbind v2016, v2020, and 1.8 in CSAR-HiQ, respectively. 
This could be because the topological graph-based model disregards crucial spatial information. 

Pafnucy, OnionNet, SchNet, DimeNet, SphereNet, and other spatial graph models leverage the 3D spatial structures of proteins and ligands. 
These models have RMSE of approximately 1.454, 1.444, and 1.715 in PDBbind v2016, v2020, and CSAR-HiQ, which is better than the topological models. 
Due to the computational complexity of these models, however, it is difficult to apply them to macromolecular systems like proteins. 
Additionally, these models make poor use of long-range information. 

The SIGN model is a comprehensive model that exploits the chemical element pair as the long-range interaction effect and extracts the angle and spatial info. 
Its RMSE of the prediction effect is 1.31 in PDBbind and 1.735 in CASE-HiQ, which is superior to the simple spatial model and topological model. 
However, like the spatial model, this model consumes a substantial amount of time and computing resources.

Finally, the results of our GLI-based models (with different chemical info modules) are better than the above SOTA models. 
For PDBbind v2016, GLI-based models are the first model to make RMSE less than 1.3.  
For CSAR-HiQ as test data, GLI-based models are the first model to make RMSE less than 1.55 and MAE less than 1.2.  
These results demonstrate that our GLI framework has encouraging performance in predicting protein-ligand affinity.

\subsubsection{Robustness to The Cutoff Distance $\mu$} \label{Cutoff analysis}
In our GLI framework, the local interaction module is directly affected by the size of the local interaction graph, which is determined by the cutoff distance $\mu$. 
Consequently, we quantify the impact of the cutoff distance $\mu$ on the performance of our framework. 

\paragraph{Local Interaction Graph} 
The local interaction graph contains the nodes close to each other between proteins and ligands. 
As shown in Figure~\ref{fig: cutoff}, the number of nodes contained in the local interaction graph increases gradually as the cutoff distance $\mu$ changes. 
When the cutoff distance is set to 1, the local interaction graph contains, on average, 0.02\% of all ligand nodes and 0.001\% of all protein nodes. 
The local interaction graph can contain all ligand nodes (99.3\%) and 20.2\% of protein nodes when the cutoff distance is 5. 
When the cutoff distance is set to 11, the local interaction graph contains the vast majority of ligand nodes (99.9\%) and protein nodes (99.9\%), with an average of 441 nodes. 
At this point, the local interaction graph can be viewed as an accumulation of proteins and ligands, with an average of 5,470 interaction edges and an extremely low computational efficiency. 

\paragraph{Results under Various Cutoff Distances} 
For GLI models (including chemical info module, global interaction model, and local interaction model), when the cutoff distance $\mu$ is less than 3, there will be no local interaction edge for some proteins and ligands, so the short-range interaction cannot be estimated at this time. 
Additionally, when the cutoff distance $\mu$ changes from 4 to 7, the effect of the model increases firstly and then decreases. 
When the cutoff distance $\mu$ is set to 5, our GLI framework can achieve optimal performance, which corresponds to the node distribution of the local interaction graph.
The results also indicate that it is inappropriate to include too many or too few short-range interaction pairs in the local interaction model. 
Too many pairs will diminish the calculation effect and increase calculation time. 
Too few pairs will make it difficult for the model to account for all short-range interactions. 
When the cutoff is set to 5, the model can utilize all ligand nodes and a subset of protein nodes. 
Consequently, the prediction result is the best possible.

\subsubsection{Ablation Study of GLI}
\begin{table*}[tb]
\centering
\begin{threeparttable}
\caption{T-test results of different modules in GLI.\tnote{a}} 
\label{tab:T-test result}
\small 
\begin{tabular}{l c c l c c l c c l}
\toprule
GLI & \multicolumn{3}{c}{PDBbind v2016} & \multicolumn{3}{c}{PDBbind v2020} & \multicolumn{3}{c}{CSAR-HiQ} \\
\cmidrule(r){2-4}  \cmidrule(r){5-7}   \cmidrule(r){8-10} 
Model &  RMSE$\downarrow$ & $\Delta$RMSE$\downarrow$ & T-test$\uparrow$  &   RMSE$\downarrow$ & $\Delta$RMSE$\downarrow$ & T-test$\uparrow$ &  RMSE$\downarrow$ & $\Delta$RMSE$\downarrow$ & T-test$\uparrow$  \\ \midrule
GLI-0-c   & 1.599  &   &  & 1.593  &     & & 1.936    &  &   \\
GLI-0-cg   & 1.399  & -0.199 & $15.616^{***}$  & 1.457 & -0.136 &  $15.728^{***}$ &  1.646 & -0.290 & $3.949^{***}$ \\
GLI-0-cgl & 1.321 & -0.077 & $5.768^{***}$   & 1.354 & -0.102 &  $10.426^{***}$  & 1.595 & -0.051 & $2.310^{*}$  \\ \midrule
GLI-1-c   & 1.554 &    &                 & 1.568 &      &    & 1.849 &   &  \\
GLI-1-cg   &1.392  & -0.162  & $12.742^{***}$   &  1.417 & -0.151   &  $19.721^{***}$  & 1.638 & -0.211 &  $5.724^{**}$   \\
GLI-1-cgl  &  1.294 & -0.098  & $6.946^{***}$  &   1.353 & -0.064 &  $5.080^{***}$     &  1.597 & -0.040 &  $2.014^{*}$  \\ \midrule
GLI-2-c  & 1.604 &   &      &  1.578  &   &    & 2.316 &  &  \\
GLI-2-cg  &  1.407 & -0.197  & $16.596^{***}$  & 1.438 & -0.141 & $16.797^{***}$  & 1.587 & -0.729 & $15.749^{***}$ \\
GLI-2-cgl   &  1.347 & -0.060  & $5.127^{***}$    & 1.405 & -0.032   & $3.582^{***}$  &1.549 & -0.038 &  $2.447^{*}$  \\ \bottomrule
\end{tabular}
  \begin{tablenotes} 
	\item[A] $^*$,  $^{**}$, $^{***}$ means the P-value is less than 5\%, 1\%, 0.1\%.
 \end{tablenotes} 
\end{threeparttable}
\end{table*}

As shown in {Table~\ref{tab:ablation result}}, we conducted ablation experiments on GLI model, with various network structures (GAT+GCN, GIN, and GCN2) as chemical embedding models, and gradually added a global interaction module and local interaction module for experiments. 
To analyze whether the changes brought by different modules are significant or just from random effects, we further make a T-test experiment for the predicted result in {Table~\ref{tab:T-test result}}. 
The T-test result includes the test result and significance level. 
The corresponding analysis is as follows:

\paragraph{Chemical Info Module} 
When only the chemical info module is included, our GLI framework is equivalent to using only biochemical information of data. 
We use (GAT+GCN), GIN, and GCN2 as the network structures to implement the chemical info module, respectively. 
The experimental results show that the predicted RMSE of PDBbind v2016, v2020 and CSAR-HiQ datasets reach 1.5-1.6 and 1.8-2.3 respectively, indicating the chemical info module is not insufficient to make an accurate prediction. 

\paragraph{Global Interaction Module} 
Based on the chemical info module, we add the global interaction module to model the long-range interaction between protein and the ligand. 
This module gathers protein and ligand as one node, respectively. 
Then it calculates the interaction effect between the protein and the ligand. 
We can observe that for GLI-0 (GAT + GCN as chemical info module), after adding global interaction, the predicted RMSE in PDBbind v2016, v2020 and CSAR-HiQ decreased by 0.199, 0.136, and 0.290.
The p-value results of the T-test are less than 0.1\%, indicating that the decrease in RMSE is statistically significant.
Besides, to test the generalizability of the global interaction module, we apply GIN and GCN2 separately as the chemical info module and add the global interaction module. 
The same results were also observed in the PDBbind v2016, v2020, and CSAR-HiQ datasets. 
Moreover, introducing the global interaction module has a small impact on the calculation time of GLI. 
These results indicate that the global interaction module is a simple and effective structure with strong generalizability. 
Specifically, the global interaction module significantly improves the accuracy of predictions by decreasing RMSE from 0.136 (8.5\%) to 0.729 (31.5\%) without increasing too much computational demand.

\paragraph{Local Interaction Module} 
Based on the preceding model, we add the local interaction module. 
This module estimates the embedding of non-covalent bonds between protein and ligand within the cutoff distance $\mu$ to represent the short-range interactions. 
The results indicate that after adding the local interaction model to GLI-1-cg model, which the contains chemical info module (GIN) and global interaction module, the predicted result in the PDBbind v2016 dataset is pretty satisfied with the RMSE of 1.294 (decreased 0.098) and the T-test result of 6.946 (P-value$<0.1\%$). 
Comparable results are also obtained in GLI-0, GLI-2 and PDBbind v2020, CSAR-HiQ datasets, demonstrating that in statistics, the local interaction module can significantly improve binding affinity prediction by reducing predicted error by 0.102 at most . 

In addition, based on GLI-0-cgl model, we add spatial graph-based models (SchNet, DimeNet) to update the node embedding of the local interaction graph. 
The comparison results show that these spatial models provide few or even negative improvements. 
This could be because the existing spatial model aims to update the node embedding, which would reduce the gap between the nodes of protein and ligand. 
Thus, the prediction results do not improve or even decline.

\subsubsection{Efficiency Analysis} 
The running time and resource consumption are vital indicators for the promotion and application of the models. 
As shown in Table~\ref{tab:experiment result0} and~\ref{tab:ablation result}, the five baseline models differ significantly in terms of prediction effect, GPU memory, and time consumption. 
In the experiment, batch size was set to 32, and epoch was set to 200. 
For the traditional machine learning models like RF-Score, have a poor prediction effect, but they do not require GPU resources, and their calculations are high-speed (0.08 hours for predicting). 
For CNN models such as Pafnucy, OnionNet, they treat molecules as images, resulting in a poor prediction effect and high GPU RAM (over 16 GB) and computing time. 
For topological (2D) graph-based models such as GAT, GCN, etc., they mainly use the topological structure and chemical info of the molecule. 
Its prediction effect is slightly better than the previous models, and its demands for memory and calculation time are modest (1,000-2,000 MB of GPU RAM and 0.1-0.5 hours for calculation). 
For spatial (3D) graph-based models such as SchNet, DimeNet, SphereNet, etc., they primarily employ the spatial information of the molecule, so their prediction effects are marginally superior to those of topology models. 
However, its resource consumption is far higher than the topology model, taking more than 10 times the GPU RAM usage and 4 times the running time later, even when the OOM (Out-of-Memory) issue occurs for DimeNet and SphereNet. 
The SIGN model makes use of spatial information and some chemical information in the molecule, and its prediction effect is also excellent. 
However, similar to the spatial graph-based models, this model still requires a significant amount of GPU RAM (nearly 20,091 MB) and running time (almost 2.73 hours). 
In contrast, in our GLI framework, with the gradual addition of three modules, the effect of the model eventually surpasses the SOTA model.
Its GPU RAM usage is only slightly higher than the topology models (less than 2,000 MB), while its running time is significantly less than the spatial models (less than 0.6 hours). 
This result demonstrates that our GLI framework has moderate efficiency in running time and resource consumption. 

\section{Conclusion}
In this paper, concentrating on the prediction of protein-ligand binding affinity, we propose a global-local interaction framework called GLI from the perspective of multi-level inter-molecular interactions in the binding process. 
GLI consists of a chemical info module, a global interaction module, and a local interaction module to learn the biochemical information, the long-range interaction and the short-range interaction effect between proteins and ligands.
In addition to superior prediction accuracy and comparable computational efficiency, the experimental results demonstrate that our framework is highly generalizable and scalable.

\paragraph{Future Work}
The prediction performance of binding affinity can be enhanced in the future by upgrading three modules of our GLI framework. 
For the chemical info module, researchers may attempt to incorporate spatial information about atoms to develop a more accurate model for extracting chemical information. 
For the global interaction module, researchers can seek to introduce an attention mechanism based on global space and distance, as well as add some node-to-node long-range interactions using a sampling algorithm.  
Besides, it's a commendable effort to incorporate principles of microphysics and chemistry into the local interaction module to raise the appropriate level of short-range interactions.

\bibliographystyle{IEEEtran}
\bibliography{main}

\end{document}